\documentclass[]{spie}  

\usepackage{amsmath,amsfonts,amssymb}
\usepackage{graphicx}
\usepackage[colorlinks=true, allcolors=blue]{hyperref}
\usepackage[utf8]{inputenc}
\usepackage{float}     
\usepackage{subfigure} 
\usepackage{eurosym}
\usepackage{color}    
\usepackage{verbatim} 
\usepackage[small,hang]{caption2}

\newcommand{\dd}{\,\mathrm{d}}

\title{MIGA: Combining laser and matter wave interferometry for mass distribution monitoring and advanced geodesy}
\author[a,b]{B. Canuel}
\author[a,b]{S. Pelisson} 
\author[a,c]{L. Amand}
\author[a,b]{A. Bertoldi}
\author[a,d]{E. Cormier}
\author[a,c]{B. Fang}
\author[a,e]{S. Gaffet}
\author[a,c]{R. Geiger}
\author[f,g]{J. Harms}
\author[a,c]{D. Holleville}
\author[a,c]{A. Landragin}
\author[a,b]{G. Lefèvre}
\author[a,d]{J. Lhermite} 
\author[a,c]{N. Mielec}
\author[h]{M. Prevedelli}
\author[a,b]{I. Riou}
\author[a,b]{P. Bouyer}

\affil[a]{MIGA Consortium}
\affil[b]{LP2N, Laboratoire Photonique, Numérique et Nanosciences, Université Bordeaux--IOGS--CNRS:UMR 5298, rue F. Mitterrand, F--33400 Talence, France.}
\affil[c]{LNE--SYRTE, Observatoire de Paris, PSL Research University, CNRS, Sorbonne Universités, UPMC Univ. Paris 06, 61 avenue de l’Observatoire, F--75014 Paris, France}
\affil[d]{CELIA, Centre Lasers Intenses et Applications, Université Bordeaux-CNRS-CEA-UMR 5107, F--33405 Talence, France}
\affil[e]{GEOAZUR, UNSA, CNRS, IRD, OCA , 250 rue Albert Einstein, 06560 Valbonne, France}
\affil[f]{INFN, Sezione di Firenze, I-50019 Sesto Fiorentino, Italy}
\affil[g]{Università degli studi di Urbino `Carlo Bo', I-61029 Urbino, Italy}
\affil[h]{Dipartimento di Fisica e Astronomia, Universit{\`a} di Bologna, Via Berti-Pichat 6/2, I--40126 Bologna, Italy}

\authorinfo{Further author information:\\ Send correspondence to benjamin.canuel@institutoptique.fr}

\begin{document} 
\maketitle

\begin{abstract}
The Matter-Wave laser Interferometer Gravitation Antenna, MIGA, will be a hybrid instrument composed of a network of atom interferometers horizontally aligned and interrogated by the resonant field of an optical cavity. This detector will provide measurements of sub Hertz variations of the gravitational strain tensor. MIGA will bring new methods for geophysics for the characterization of spatial and temporal variations of the local gravity field and will also be a demonstrator for future low frequency Gravitational Wave (GW) detections. MIGA will enable a better understanding of the coupling at low frequency between these different signals. The detector will be installed underground in Rustrel (FR), at the ``Laboratoire Souterrain Bas Bruit" (LSBB), a facility with exceptionally low environmental noise and located far away from major sources of anthropogenic disturbances. We give in this paper an overview of the operating mode and status of the instrument before detailing simulations of the gravitational background noise at the MIGA installation site. 

\end{abstract}


\section{INTRODUCTION}

After the development in the 1980's of cooling and trapping techniques for neutral atoms \cite{Chu1997}, it became possible to demonstrate the wave behavior of massive particles at the atomic level \cite{Berman1997}, using micro-fabricated gratings \cite{Carnal91,Keith1991} or optical lattices \cite{Riehle91,Kasevich91} to coherently manipulate the atomic wave function. Since then, the field of atom interferometry has known an incessant development \cite{Barrett2016}, and nowadays sensors based on matter wave interference phenomena are used to probe inertial forces (like acceleration \cite{Peters2001}, gravity gradient \cite{Snadden98}, rotation \cite{Gustavson1997,Canuel2006}), measure fundamental constants (like the gravitational constant G \cite{Fixler2007,Rosi2014}, and $h$/M \cite{Bouchendira2011}), and are investigated as possible means to test a long list of disparate effects and theories (like the weak equivalence principle in general relativity \cite{Fray2004,Schlippert2014,Barrett2015,Zhou2015,Bonnin2015,Aguilera2014,Hartwig2015}, matter neutrality \cite{Arvanitaki2008}, and dark energy \cite{Burrage2015,Hamilton2015b}). 

A few years ago large scale atom gradiometers have been proposed to monitor the strain tensor in the infra-sound bandwidth \cite{Dimopoulos2009}, with the aim of detecting GWs. Different configurations have been considered to install very long baseline atom interferometers in space \cite{Hogan2011,Yu2011,Graham2013,Hogan2015,Chiow2015} and on ground \cite{Harms2013}. We are building a large scale atom interferometer array in an underground laboratory with the two-fold aim to measure tiny variations of gravity induced by geophysical phenomena and to implement a demonstrator for sub-Hz GW detection, with an initial strain sensitivity of the order of 10$^{-13}$. Future evolutions of this instrument could enable to observe on Earth GW sources in a frequency band forbidden to optical detectors. This prospect assumes a new relevance in relation to the recent first detection of GWs \cite{Abbott2016}, which shifts the interest to GW astronomy and hence to the widening of the observation bandwidth through different classes of detectors \cite{Moore2014}.

\section{MEASURING STRAIN VARIATIONS WITH AN IN-CAVITY ARRAY OF ATOM INTERFEROMETERS}
The MIGA antenna \cite{Canuel2014, Geiger2015}  will consist of three $^{87}$Rb Atom Interferometers (AIs) horizontally aligned and interrogated by the resonant field of a 300 m long cavity. By using a three pulse $\pi/2$, $\pi$, $\pi/2$ sequence\cite{Kasevich91}, each AI becomes sensitive to both horizontal accelerations and to variations of the phase of the interrogation field during the interrogation sequence. In a gradiometer configuration, couples of in-cavity interferometers will therefore provide access to the horizontal gravity gradient and more generically to the differential phase variation between the distant atom sources. Such differential phase fluctuations may arise from strain variations of the space-time metric induced by GWs\cite{Dimopoulos2008}. 

\subsection{MIGA Atom Interferometers}
\begin{figure}[h!]\centering
\includegraphics[width=.5\paperwidth]{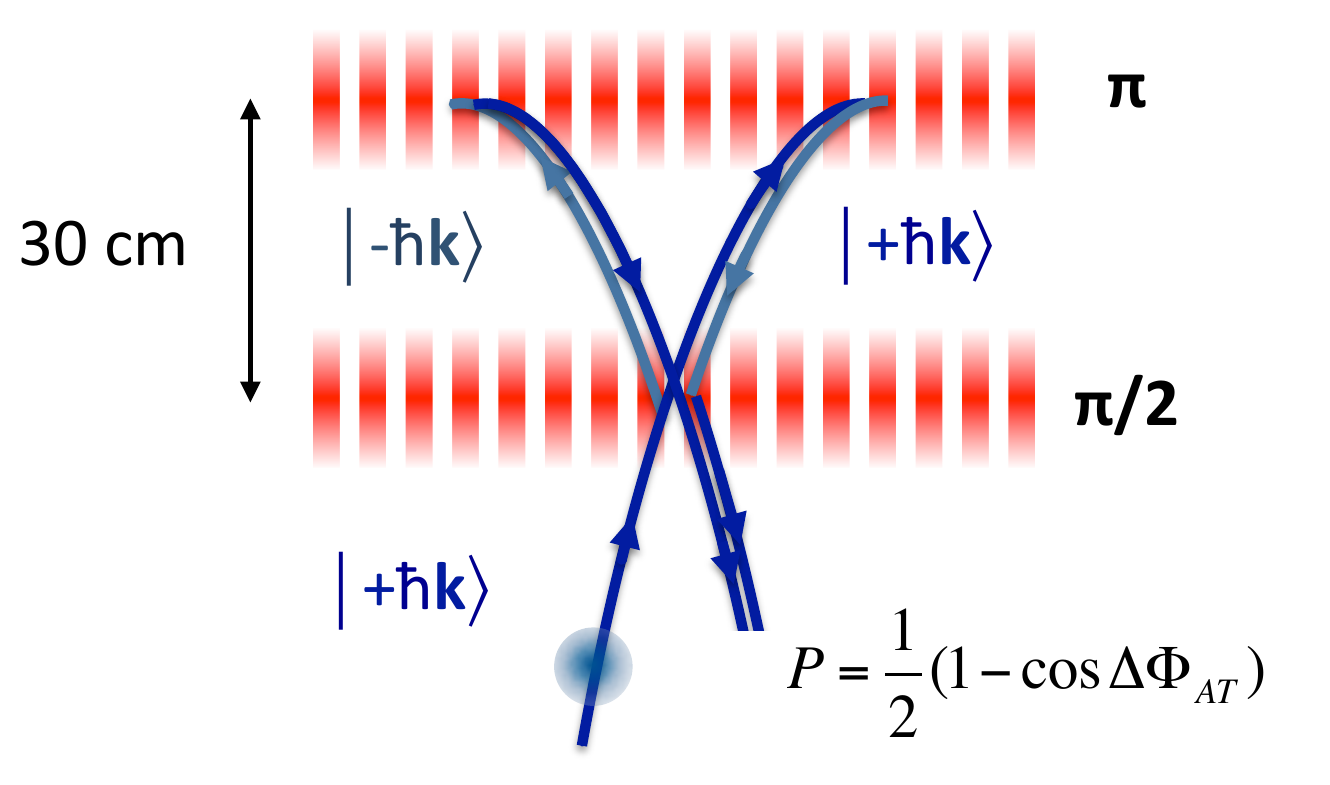}
\caption{\label{MZITF}Geometry of the MIGA atom interferometers. Atoms are launched on the vertical trajectory and experience a set of cavity enhanced $\pi/2$, $\pi$, $\pi/2$ pulses that creates a matter-wave interferometer on the external states $\left |+\hbar k\right\rangle$ and $\left |-\hbar k\right\rangle$. The $\pi$ pulse is realized at the apex of the atom trajectories.}
\end{figure}

The different AIs of the antenna are simultaneously created using a set of three in-cavity  $\pi/2$, $\pi$, $\pi/2$ light pulses. The matter-waves are manipulated using Bragg diffraction \cite{Martin88} of the atoms on the cavity standing wave. Such process couples atomic states of momentum $\left |+\hbar k\right\rangle$ and $\left |-\hbar k\right\rangle$ where $k$ is the wave vector of the interrogation field. The interferometer geometry is described in Fig. \ref{MZITF}.
The atoms, initially in the $\left |+\hbar k\right\rangle$ state, first experience a $\pi/2$ pulse creating an equiprobable coherent superposition between the $\left |+\hbar k\right\rangle$ and $\left |-\hbar k\right\rangle$ states. The matter-waves are then deflected by the use of a $\pi$ pulse reversing the atomic states, before being recombined with a second $\pi/2$ pulse. At the interferometer output the transition probability P between the states is given by:
\begin{equation}
\label{2Waves}
P=\frac{1}{2}(1-\cos \Delta \phi_{AT}).
\end{equation}

\subsection{Strain measurement with the Atom Interferometers of the antenna}

The atom phase shift $ \Delta\phi_{AT}(X_i)$ of the AI located at $X_i$ along the cavity will measure the horizontal accelerations $s_a(X_i)$ together with strain variations induced by GWs. We consider in the following that fluctuations of cavity mirror position x$_1$(t) and x$_2$(t) and laser frequency noise $\delta\nu$(t) are the only sources of experimental noise\footnote{Other effects such as wavefront aberration\cite{Hogan2011} or input beam jitter \cite{Sorrentino2014} should also be carefully evaluated.}. Considering variations of these effects with characteristic frequencies smaller than the linewidth of the interrogation cavity, the atom phase shift for an AI placed in the cavity is similar to that of a simple retro-reflection configuration \cite{Chaibi2016}:
\begin{equation}
\label{AIresponse}
\Delta\phi_{AT}(X_i)=\frac{4\pi\nu_0}{c}s_{x_{2}}+\frac{4\pi}{c}\left[-s_{\delta\nu}+\frac{\nu_0}{2}s_h\right]\left(X_i-L\right)+s_a(X_i)+\epsilon (X_i),
\end{equation}
where $\nu_0$ is the laser frequency, $h$ the GW strain variation, L the cavity length and $\epsilon (X_i)$ is the detection noise (atom shot noise). The term $s_a(X_i)$ accounts for the weighting of the time-dependent fluctuations of local accelerations by the sensitivity function of the AI to accelerations: $s_a(X_i)=\int^{\infty}_{-\infty}s_a(t)\frac{da(X_i,t)}{dt}dt$. Other time-fluctuating effects are weighted by the regular phase sensitivity function of the AI \cite{Cheinet08} ($s_{x_{2}}$,$s_{\delta\nu}$ and $s_h$). The differential signal between two in-cavity AIs placed at $X_i$ and $X_j$ is therefore:
\begin{equation}
\label{AIresponse2}
\Delta\phi_{AT}(X_i)-\Delta\phi_{AT}(X_j)=\frac{4\pi}{c}\left[-s_{\delta\nu}+\frac{\nu_0}{2}s_h\right]\left(X_i-X_j\right)+s_a(X_i)-s_a(X_j)+\epsilon (X_i)-\epsilon (X_j).
\end{equation}
This signal thus presents strong immunity to vibration of the cavity end mirror. Vibrations of other optical elements may also re-couple to differential measurement though generation of frequency noise\cite{LeGouet2007} $\delta\nu$(t), but this contribution will remain negligible with respect to detection noise in the initial instrument configuration.  

This gradiometer configuration couples gravity gradient signals with strain variations induced by GW over a certain baseline. The use of an array of correlated AIs offers simultaneous measurements of both effects over different characteristic lengths.
Such configuration may enable to reduce the influence of gravity gradient fluctuations on the GW measurements by taking advantage of the correlation properties of the local gravity acceleration noise \cite{Chaibi2016}, and opens the possibility to detect gravitational radiation at low frequency on Earth. The use of a dense network of AIs may also permit to measure the space-time variations of the local gravity field, with important outputs in geophysics for mass transport monitoring, hydrogeology and underground survey.

In Sec. \ref{NN}, we detail the first projection of background gravity gradient noise (called in the following Newtonian Noise, NN) on the differential signals of the MIGA AIs. Remarkably, the precise modelization of such noise is at the heart of NN rejection methods for GW detection\cite{Chaibi2016}.

\section{THE MIGA ANTENNA}

In this section, we give an overview of the antenna operating principle, focusing on the AIs and the interrogation cavity. We also describe the infrastructure work required at ``Laboratoire Souterrain Bas Bruit" (LSBB) to host the MIGA instrument.

\subsection{Instrument}

\subsubsection{Atom Interferometers}\label{AI}
\begin{figure}[htp]\centering
\includegraphics[width=.7\paperwidth]{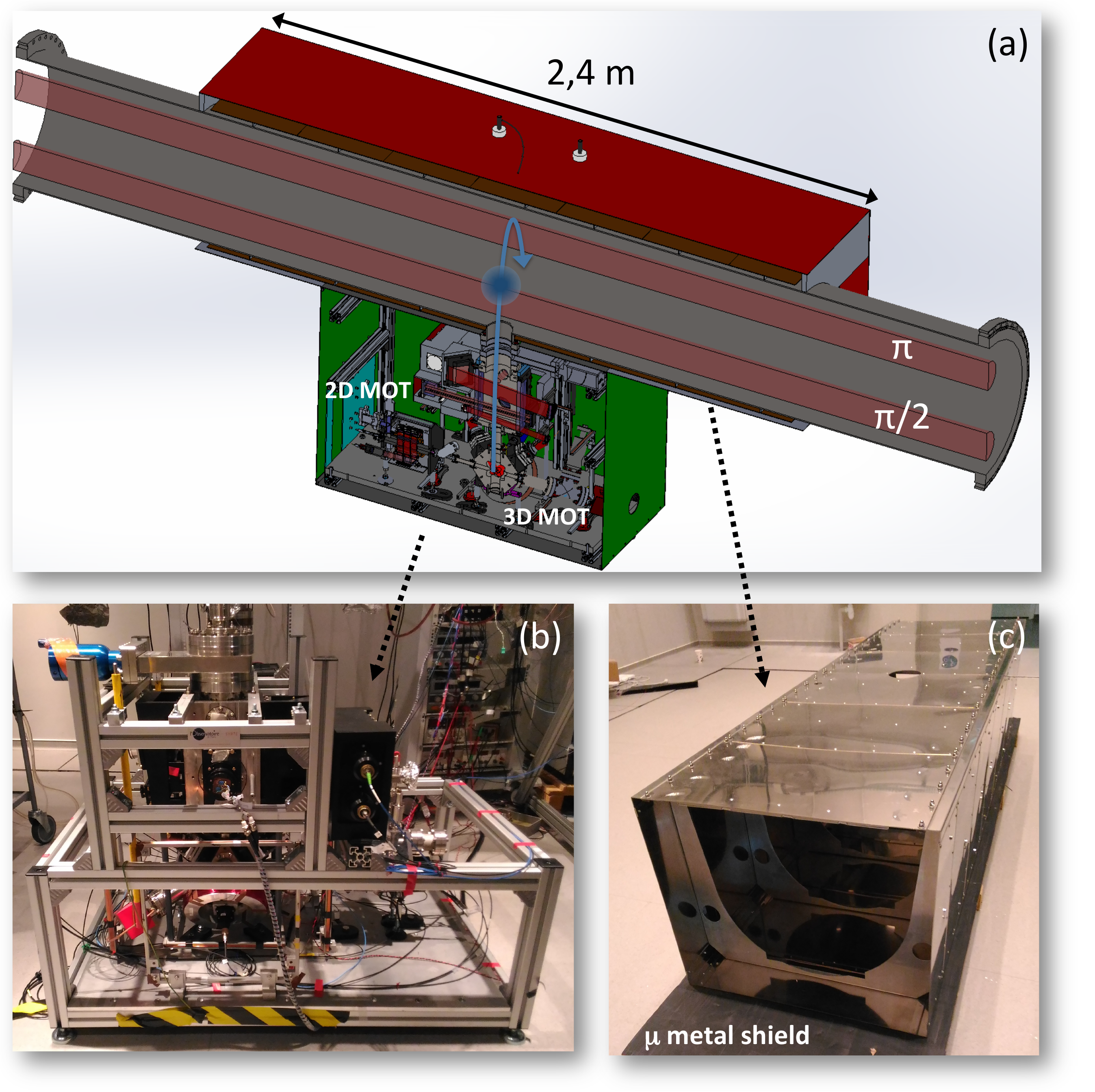}
\caption{\label{MigaAI}Description of the MIGA atom interferometers. (a) View of the AI including 2D-MOT, 3D-MOT, preparation and detection systems. A set of 3 different $\mu$metal shields (red, brown, green colors) are used to screen the system from magnetic fields. (b) Photography of the 2D-MOT, 3D-MOT, preparation and detection systems. (c) Photography of the interior $\mu$metal shield, placed along the vacuum system of the interrogation cavity (brown shield of (a)).}
\end{figure}

The atom sources of MIGA consist in $^{87}$Rb clouds cooled down to the sub-Doppler regime in a 3D Magneto Optical Trap (3D-MOT) loaded by a 2D Magneto Optical Trap (see Fig.\ref{MigaAI}). 
The clouds are vertically launched at a controlled velocity of $\approx$4 m/s by shifting the relative frequency between the top and bottom lasers of the 3D-MOT. Before entering the interrogation region, the quantum state of the sources is prepared using a set of Raman pulses. A first velocity-selective pulse prepares the atoms in the $m_F$=0 Zeeman sub-level of the $F=1$ hyperfine fundamental state with a longitudinal velocity distribution corresponding to a temperature of few hundreds of nano Kelvins, while the remaining atoms on the $F=2$ state are blasted using a laser resonant on the cooling transition. This sequence is repeated with a shorter Raman pulse to remove atoms on $m_F\neq 0$ produced by spontaneous emission at the first pulse.

At the apex of their trajectory, the clouds experience a set of in-cavity $\pi/2$, $\pi$, $\pi/2$ pulses before returning in the detection region where the transition probability is measured. Here, we first transfer coherently the population of the $\left |F=2,-\hbar k\right\rangle$ state towards $\left |F=1,+\hbar k\right\rangle$ using a Raman transition and then use a fluorescence detection on the internal states of the atoms.
The fluorescence signal is used to reconstruct the normalized atomic populations of the two internal states and then to calculate the transition probability inside the interferometer, related to the atom phase shift $\Delta\phi_{AT}(X_i)$ (see Equ.\ref{2Waves}).

\subsubsection{Laser and interrogation Cavity}\label{Cavity}
The frequency of the interrogation laser corresponds to the D2 transition of $^{87}$Rb at 780 nm. This wavelength is obtained by frequency doubling of a telecom laser at 1560 nm. The 780 nm light has to be pulsed to construct the AI. In order to have it resonant to the cavity, the fundamental radiation at 1560 nm is used to lock the resonator.

\begin{figure}[htp]\centering
\includegraphics[width=.7\paperwidth]{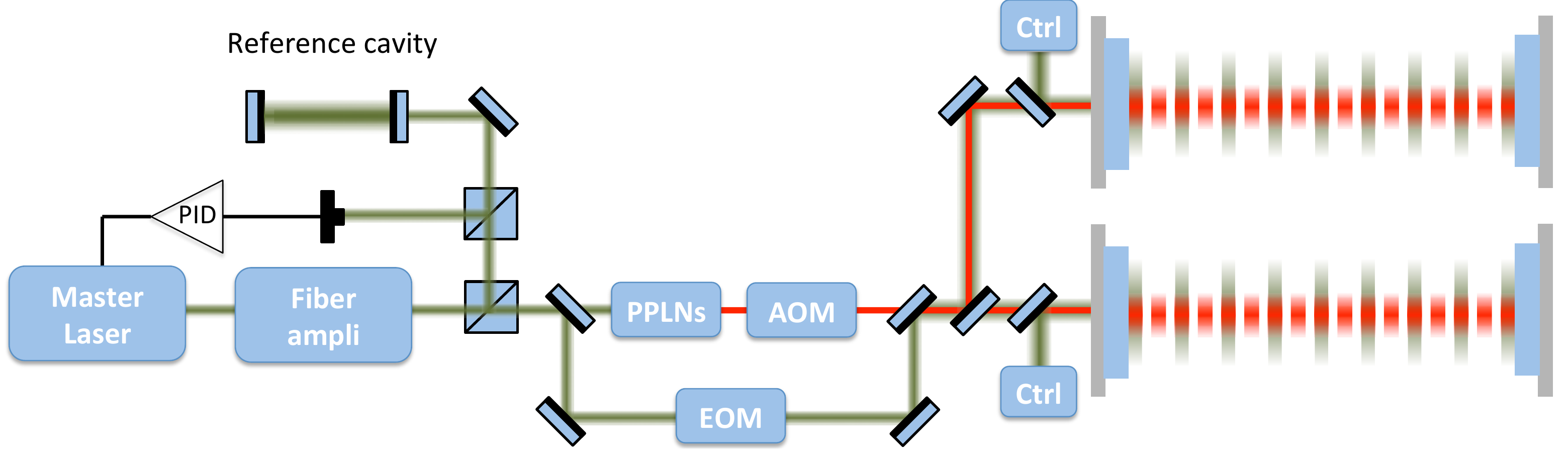}
\caption{\label{Laser_SYS} Scheme of the laser and interrogation cavity systems. green/red: 1560/780 nm radiations}
\end{figure}

The scheme of the laser and interrogation cavity system is detailed in Fig. \ref{Laser_SYS}. It consists in a master 1560 nm oscillator pre-stabilized after amplification at the $d\nu/\nu=10^{-15}$ Hz$^{-1/2}$ level using a highly stable reference cavity. Most of the 1560 nm radiation goes to a doubling stage using Periodically Pole Lithium Niobate (PPLN) crystals \cite{Chiow2012, Sane12}. The target of the setup is to obtain a few tens of Watts after doubling, to make feasible high order Bragg diffraction \cite{Giltner95} in the interrogation cavity to improve in the future the measurement sensitivity.
The 780 nm radiation passes through an Acousto-Optic Modulator (AOM) to generate the interrogation pulses. Part of the 1560 nm radiation goes through an Electro-Optic Modulator (EOM) before recombination with the 780 nm light. The beam with 1560 and 780 nm radiation is then split and injected in the two interrogation cavities.
The reflection of the two cavities at 1560 nm is used for longitudinal and angular control of the resonators.

\subsection{Site infrastructures}
The LSBB low noise underground laboratory \cite{LSBB} is a European interdisciplinary laboratory for science and technology created from the re-conversion of a former launching control system of nuclear missiles. This laboratory, located in Rustrel (FR), benefits from very low environmental noise and is located far away from major sources of anthropogenic disturbances. Thanks to its very low background noise, the LSBB is an ideal facility for site studies of next generation GW detectors and more generally for the improvement of low-frequency sensitivity of existing and future GW antennas. In such environment, MIGA aims at studying Newtonian Noise and testing advanced detector geometries for its cancellation.

The LSBB now covers 0.54 km$^2$ in surface area and consists in 4km-long horizontal drifts at a depth ranging from 0 m to 518 m, with north-south and east-west as its main orientations. The galleries give access to underground wells, vaults and voids, a clean room, and an electromagnetically shielded vault. 

\begin{figure}[h!]\centering
\includegraphics[width=.8\paperwidth]{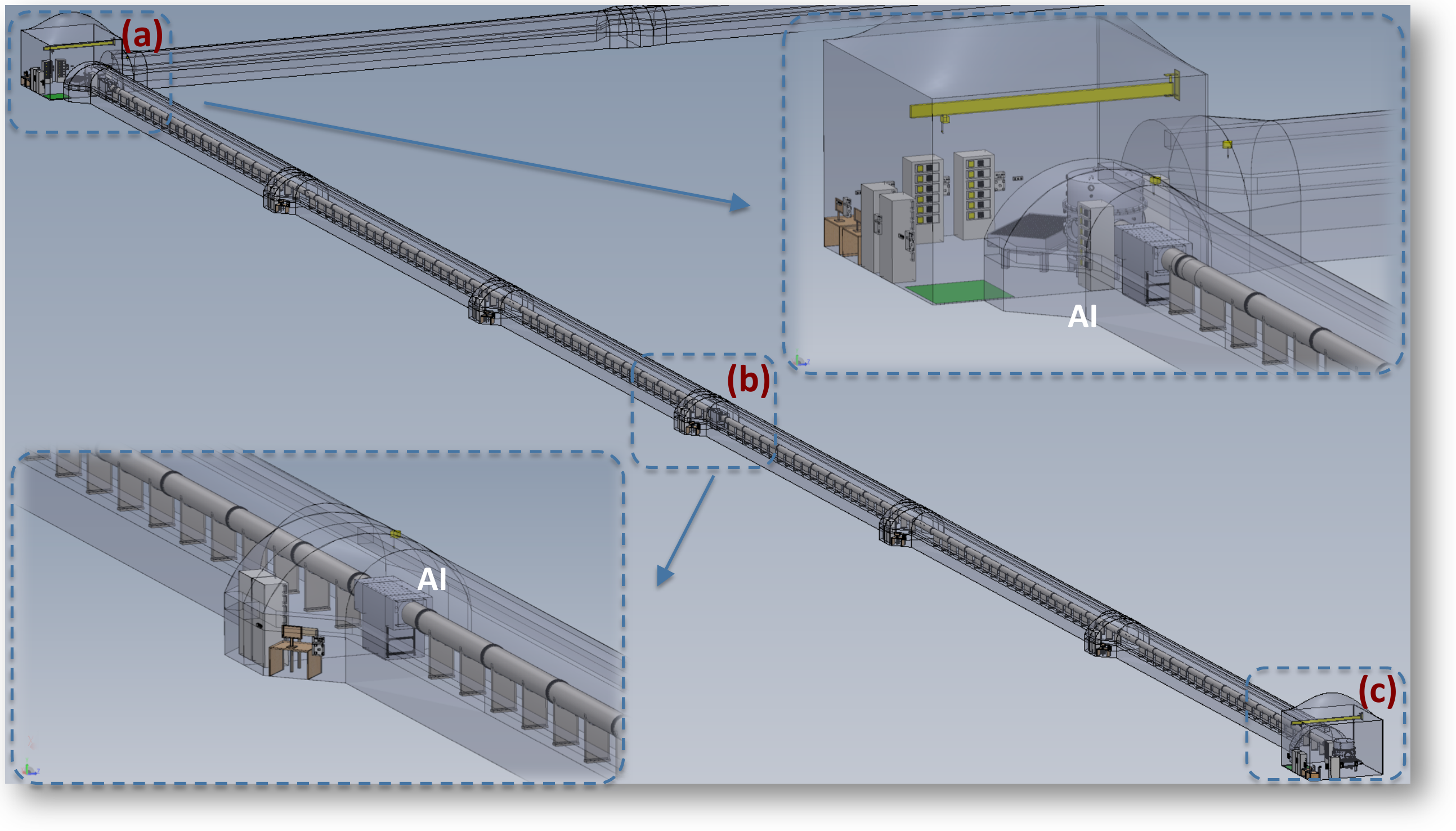}
\caption{\label{INFRALSBB}Design of the galleries dedicated to the MIGA experiment at LSBB. The AIs of the initial antenna will be located at (a), (b) and (c). Optical setups for cavity injection will be hosted in room (a).}
\end{figure}

A dedicated extension of this facility is planned to host the MIGA experiment. It consists in two perpendicular galleries 300 m long, 3.2 m width, including large cavities at their ends and regular widenings for instrumentation. Such infrastructure will host the vacuum vessel, AIs, cavity injection optical setups and various electronic and data acquisition equipments. The gallery hosting the antenna is described in Fig. \ref{INFRALSBB}: the antenna will be initially equipped with three AIs equally spaced along the gallery (locations (a),(b),(c) in Fig.\ref{INFRALSBB})

The infrastructure design is compatible with further upgrades. A second 300 m perpendicular gallery will be available to implement in the future a 2D antenna geometry. Indeed, both galleries will have widenings every 50 m to increase the number of AIs and improve the spatial resolution of the experiment.

\section{MODELIZATION OF SPATIAL AND TEMPORAL VARIATIONS OF THE GRAVITY FIELD IN AN UNDERGROUND ENVIRONMENT}\label{NN}
We report in the following the simulation of terrestrial gravity background noise, also called NN, in the underground environment of the LSBB. We evaluate the impact of NN on the strain measurement of an underground gradiometer placed at a depth $h$=250 m below the surface and using a baseline $L$=300 m (the largest available with the MIGA antenna). The NN is a gravity noise measured by the detector due to density fluctuations around the test masses. At low frequencies, the two main sources of NN are coming from the seismic environment of the detector and the atmospheric variations of air pressure \cite{Saulson84}. 
\subsection{Newtonian Noise of seismic origin}
\begin{figure}
\centering\includegraphics[width=.2\paperwidth]{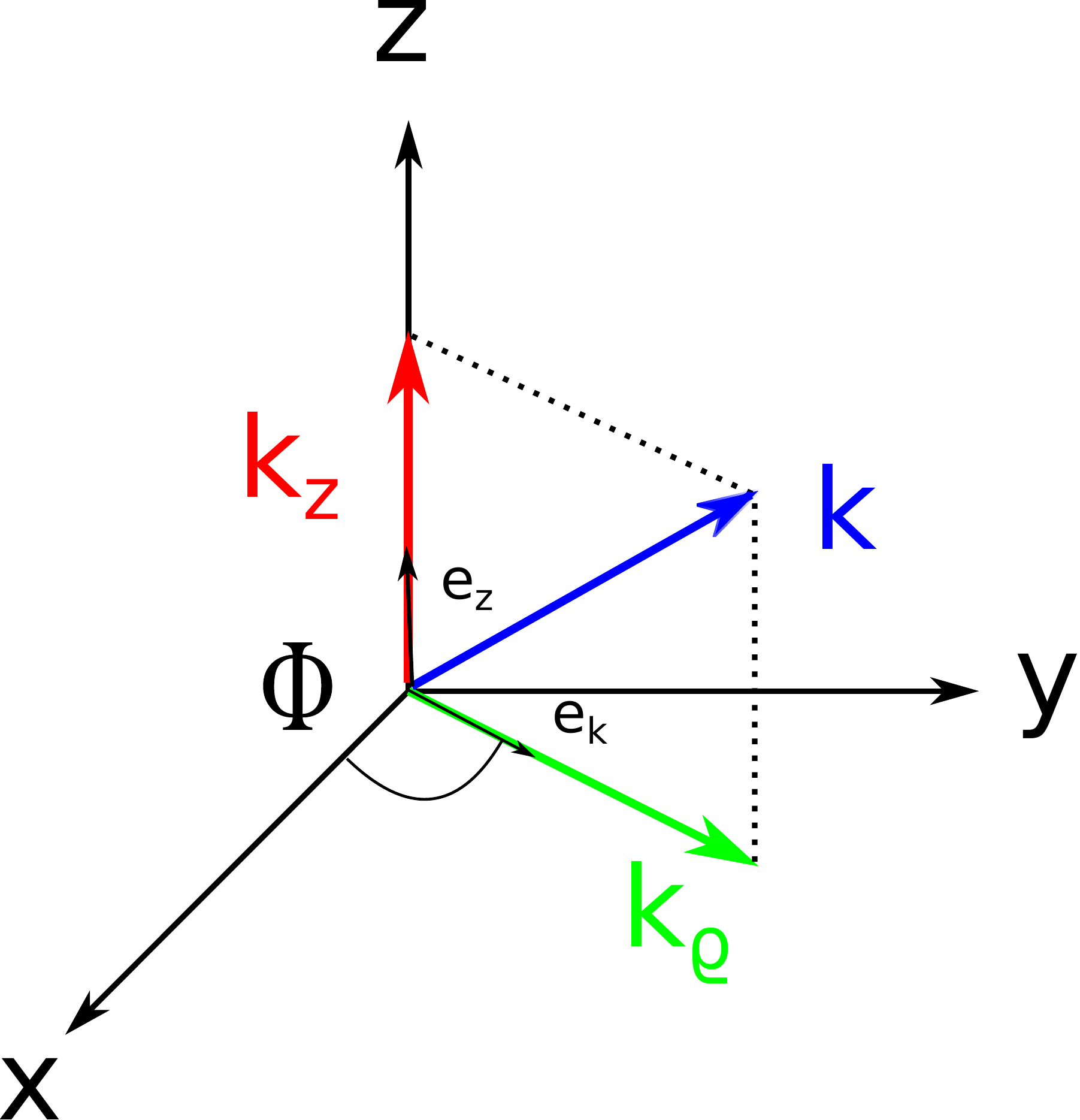}
\caption{\label{proj} Scheme of the system of coordinates}
\end{figure}

We first consider the gravity noise coming from the displacement of masses around the detector due to the seismic environment. The seismic field is mainly composed of 3 different kinds of waves \cite{Aki09}:
\begin{itemize}
\item Compressional waves which produce displacement along the direction of propagation. They are also called ``P-waves'' for {\it primary waves} as they are the first to arrive after an earthquake.
\item Shear waves which produce transverse displacement and do not exist in media with vanishing shear modulus. They are known as ``S-waves'' for {\it secondary waves}. Shear and compressional waves are referred as body-waves since they can propagate through the media in all directions.
\item Surface waves which include Love and Rayleigh wavetrains for teleseismic distances. If we consider a homogeneous medium then only Rayleigh waves can propagate on its surface and below few tens of a hertz, these waves are known to be the dominant contribution to the seismic noise \cite{Bonnefoy06}.
\end{itemize}
As MIGA is an underground detector sensitive only to frequencies under a few Hertz, we can consider in first approximation only Rayleigh waves as a source for seismic noise in our system. For simplicity, we consider the case of Rayleigh waves propagating on a homogeneous half space along a horizontal direction $\vec{e}_k$. The direction normal to the surface corresponds to the $z$-axis of the coordinate system as shown in Fig. \ref{proj}. 
A wavevector $\vec{k}$ can be split into its vertical $\vec{k}_z$ and horizontal components $\vec{k}_\varrho$ so that the three-dimensional displacement field of a Rayleigh wave reads
\begin{equation}
\vec{\xi}(\vec{r},t)=\xi_k(\vec{r},t)\vec{e}_k+\xi_z(\vec{r},t)\vec{e}_z.
\end{equation}
This displacement field induces a fluctuation of the density of the medium around the test mass of the detector. This density fluctuation creates a gravity perturbation responsible for the so-called seismic NN on the test mass. To calculate this perturbation, we have to express the continuity equation on the density fluctuation which can be written in the form
\begin{equation}
\delta\rho(\vec{r},t)=-\nabla\cdot\left(\rho(\vec{r})\vec{\xi}(\vec{r},t)\right),
\end{equation}
where we assumed that the seismic density perturbations are much smaller than the unperturbed density $\delta\rho(\vec{r},t)\ll\rho(\vec{r})$, so that self-induced seismic scattering is insignificant. The corresponding perturbation of local acceleration can be written as
\begin{equation}\begin{split}
\delta\vec{a}(\vec{r}_0,t)&= -G\int\dd V\nabla_0\frac{\delta\rho(\vec{r},t)}{|\vec{r}-\vec{r_0}|}\\
&=-G\int\dd V\rho(\vec{r})\left(\vec{\xi}(\vec{r},t)\cdot\nabla_0\right)\frac{\vec{r}-\vec{r}_0}{|\vec{r}-\vec{r}_0|^3}\\
&=G\int\dd V\rho(\vec{r})\frac{1}{|\vec{r}-\vec{r}_0|^3}\left(\vec{\xi}(\vec{r},t)-3(\vec{e}_{rr_0}\cdot\vec{\xi}(\vec{r},t))\vec{e}_{rr_0}\right),
\label{acc}\end{split}\end{equation}
with $\vec{e}_{rr_0}\equiv (\vec{r}-\vec{r}_0)/|\vec{r}-\vec{r}_0|$ and $\nabla_0$ denotes the gradient operation with respect to $\vec{r}_0$. To estimate the NN on the detector, we calculated the explicit solution of the integral \eqref{acc} for plane seismic waves of the form $\vec{\xi}(\vec{r},t)=\xi_0e^{i(\vec{k}\cdot\vec{r}-\omega t)}\vec{e}_k$. This calculation gives in cartesian coordinates
\begin{equation}
\delta\vec{a}(\vec{r}_0,t)=2\pi G \rho_0\xi_ze^{-hk_\varrho}e^{i(\vec{k}_\varrho\cdot\vec{\varrho}_0-\omega t)}\gamma
\begin{pmatrix}
i\cos\phi\\i\sin\phi\\-1
\end{pmatrix},
\end{equation}
where $\phi$ is the angle of propagation with respect to the $x$-axis and $\gamma$ is a dimensionless factor depending on the elastic properties of the half-space \cite{HarmsLivRevRel15}. The noise density of differential acceleration along a baseline of length $L$ parallel to the $x$-axis reads
\begin{equation}
S\left(\delta\vec{a}(L\vec{e}_x)-\delta\vec{a}(\vec{0});\omega\right)=\left(2\pi G\rho_0e^{-hk_\varrho}\gamma\right)^2S(\xi_z;\omega)
\begin{pmatrix}
1-2J_0(k_\varrho L)+2J_1(k_\varrho L)/(k_\varrho L)\\
1-2J_1(k_\varrho L)/(k_\varrho L)\\
2-2J_0(k_\varrho L)
\end{pmatrix},
\end{equation}
where $J_n(x)$ represents the Bessel function of the first kind and $S(\xi_z;\omega)$ is the power spectral density (PSD) of the seismic field. The seismic field is obtained by onsite measurement realized with a STS-2 sensor installed in the proximity of the future gallery hosting the detector.
The parameter $\gamma$ is obtained as a function of the speed of P- and S-waves in the medium.

\begin{figure} [h!]
   \begin{center}
   \begin{tabular}{c} 
   \subfigure
   {\raisebox{-1\height}{\includegraphics[width=.6\paperwidth]{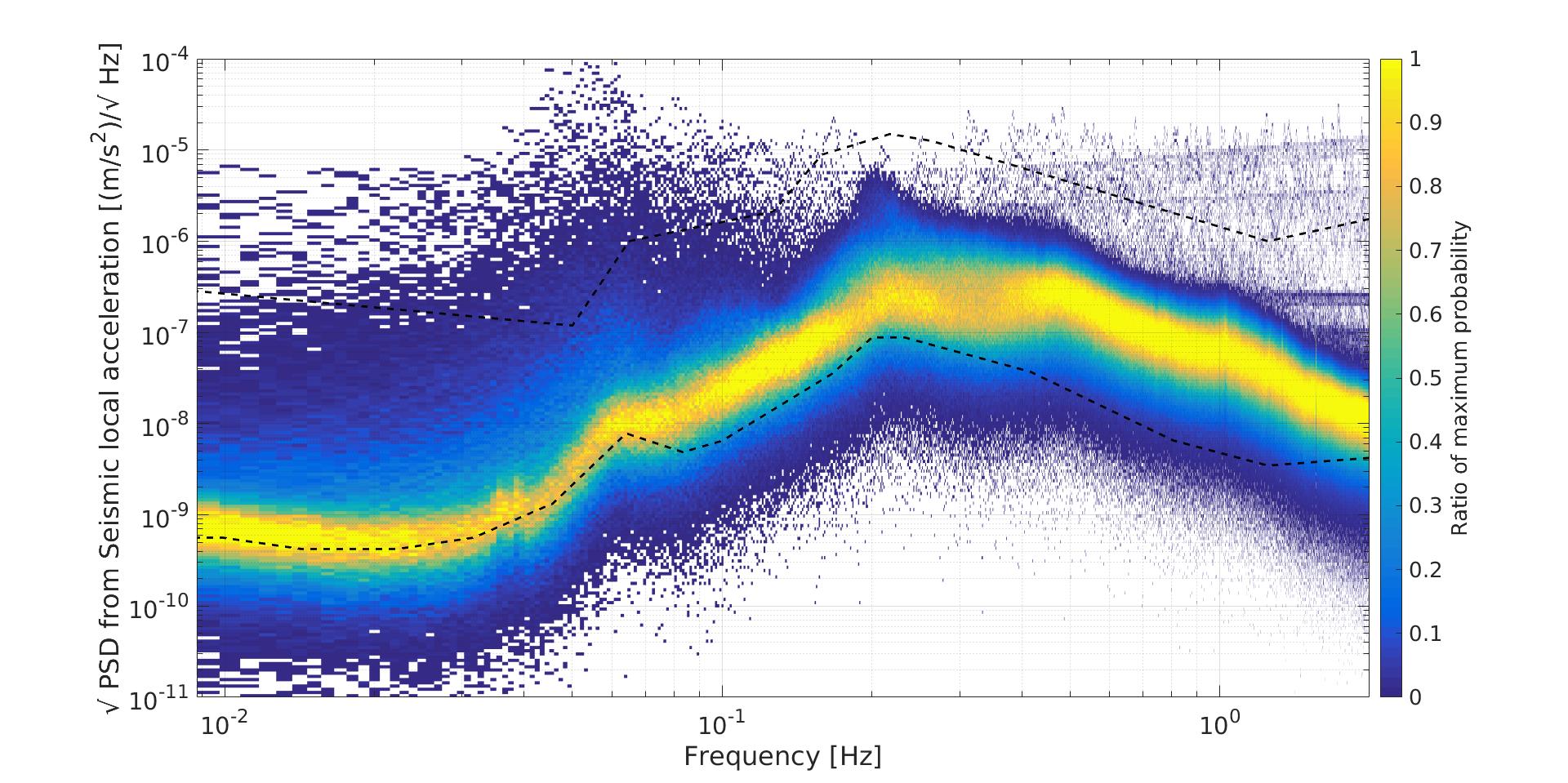}}}\\
    \subfigure{\includegraphics[width=.6\paperwidth]{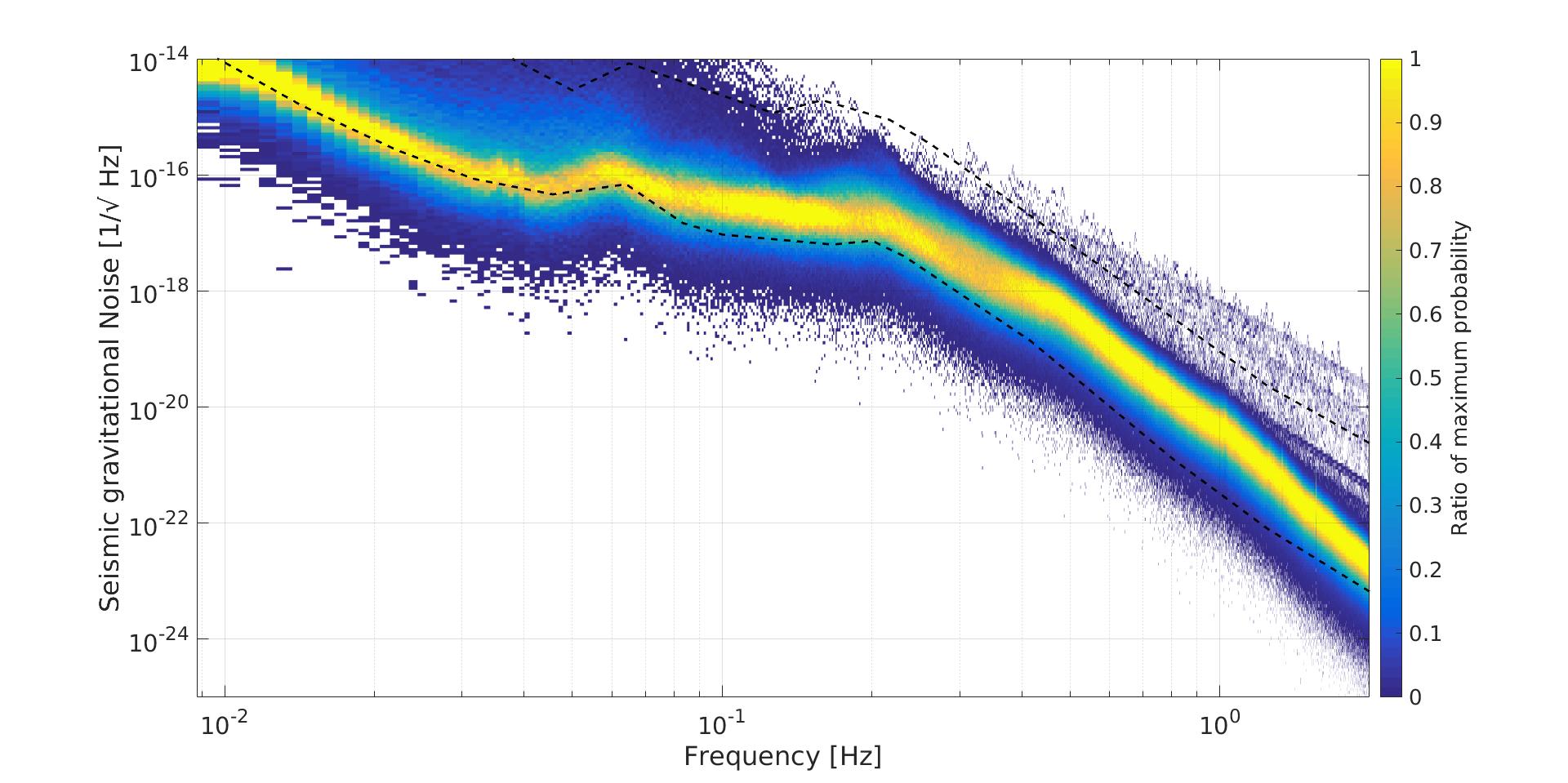}}
   \end{tabular}
   \end{center}
   \caption[example] 
   {\label{NNseis}Histograms of seismic spectrum at LSBB (Top) and modeled Rayleigh-wave Newtonian Noise (Bottom) as a function of the frequency. The dashed black lines represent the global seismic low-noise and high-noise models.}
   \end{figure}

Figure \ref{NNseis} presents the seismic spectrum recorded at LSBB and the corresponding projection of NN in terms of strain measurement obtained taking into account the baseline $L$ of the detector. We observe that this NN source limits the strain sensitivity at the level of 10$^{-16}$ Hz$^{-1/2}$ at 0.1 Hz, a strong limitation for observation of GW source in this band. This justifies further studies for NN cancellation techniques using advanced detector geometries as introduced in [\citenum{Chaibi2016}].

\subsection{Atmospheric Noise}
The second source of noise that has to be taken into account is the contribution of the atmosphere, and more precisely the contribution due to atmospheric infrasound waves coming from the pressure fluctuations.
\subsubsection{Pressure fluctuations}
Sound waves are typically understood as propagating perturbations of the atmosphere's mean pressure $p_0$. The pressure can be translated into perturbations of the mean density $\rho_0$ via the adiabatic index $\gamma\simeq1.4$ of air
\begin{equation}
\gamma\frac{\delta\rho(\vec{r},t)}{\rho_0}=\frac{\delta p(\vec{r},t)}{p_0}.
\end{equation}
The gravity acceleration of a single test mass at $h=0$ due to an infrasound wave is given by \cite{Creighton08}
\begin{equation}
\delta a_x(\vec{\varrho}_0,\omega)=-4\pi i\frac{G\rho_0}{\gamma p_0}e^{-i\vec{k}_\varrho\cdot\vec{\varrho}_0}e^{-hk_\varrho}\frac{\vec{e}_x\cdot\vec{k}}{k^2}\delta p(\omega).
\end{equation}
Averaging over all propagation directions, the strain noise measured between two test masses separated by a distance $L$ along $\vec{e}_x$ reads
\begin{equation}
S(h;\omega)=\frac{2}{3}\left(\frac{4\pi}{kL\omega^2}\frac{G\rho_0}{\gamma p_0}e^{-hk_\varrho}\right)^2S(\delta p;\omega)\left(1-j_0(kL)+2j_2(kL)\right)
\end{equation}
with $j_n(x)=\sqrt{\pi/(2x)}J_{n+1/2}(x)$ represent the spherical Bessel function.

\begin{figure}[h!]\centering
\includegraphics[width=.65\paperwidth]{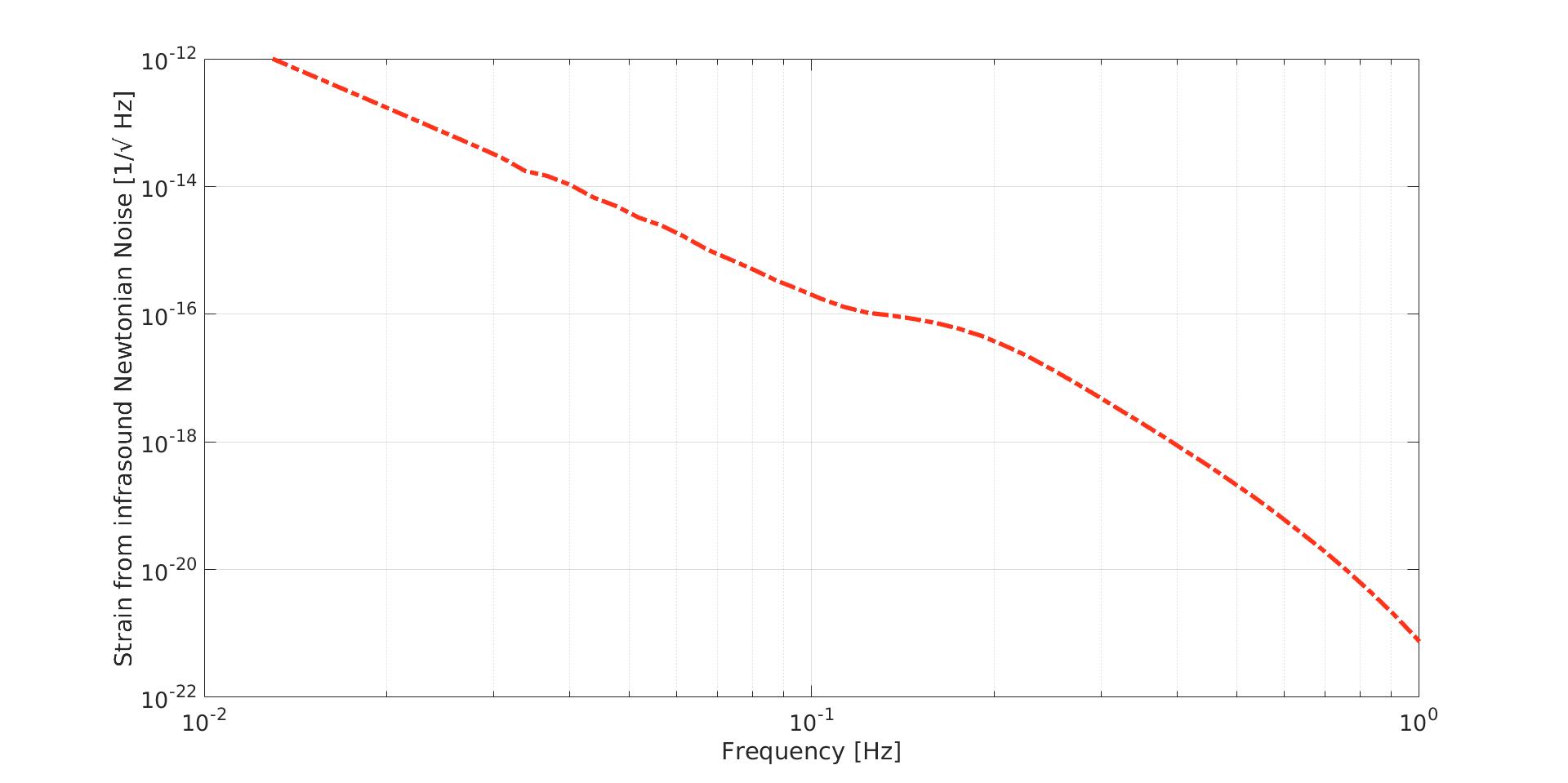}
\caption{\label{NNatm}Projection of strain noise due to atmospheric pressure fluctuations for an underground detector at 250\,m below the surface.}
\end{figure}

Figure \ref{NNatm} shows the projection of strain noise due to pressure fluctuations using mean infrasound spectra extracted from [\citenum{BowmanGRL05}]. These spectra are obtained by ambient infrasound measurements in the frequency band from 0.03 to 7 Hz realized between January 2003 and January 2004 on 21 sites all around the world.
This first projection shows that seismic and atmospheric NN sources have comparable influence on strain measurements in the band 0.1-1 Hz. Further studies are foreseen to refine the atmospheric NN projection by carrying out a pressure measurement campaign at LSBB.

\section{CONCLUSION}
The MIGA instrument will implement an array of in-cavity AIs to measure precisely the space-time variations of the local gravity field. Such measurements can be used to monitor subsurface mass transfers with potential applications in hydrogeology or underground survey. MIGA will also bring a better understanding of gravity background noise, or Newtonian Noise, that is one of the main limitations of future low frequency Gravitational Wave detectors. MIGA will also implement a demonstrator of a sub-Hz GW detector based on Atom Interferometry.

In this paper, we presented the first projection of Newtonian Noise on the strain sensitivity of MIGA gradiometers and showed limitations of strain measurements at the level of 10$^{-16}$ Hz$^{-1/2}$ at 0.1 Hz. Precise knowledge of such noise is mandatory for determining advanced detector geometries and measurement methods for NN rejection and low frequency GW detection. Indeed, the MIGA project aims to be the very first step towards the realization of a future European infrastructure for GW observation based on advanced quantum technologies.


\acknowledgments 
This work was realized with the financial support of the French State through the ``Agence Nationale de la Recherche" (ANR) in the frame of the ``Investissement d'avenir" programs: IdEx Bordeaux - LAPHIA (ANR-10-IDEX-03-02) and Equipex MIGA (ANR-11-EQPX-0028). This work was also supported by Conseil Régional d’Aquitaine (project Alisios contract number 20131603010) and by the city of Paris (Emergence project HSENS-MWGRAV). Author G. L. thanks DGA for financial support. 
\bibliography{MIGA} 
\bibliographystyle{spiebib} 

\end{document}